\documentclass[aps,pra,superscriptaddress,preprint,showpacs]{revtex4}

\usepackage{graphicx}
\usepackage{graphics}
\usepackage{amssymb}
\usepackage{amsmath}
\usepackage{epsfig}
\usepackage{latexsym}
\usepackage{color}
\usepackage{rotating}
\usepackage{subfigure}

\begin{document}

\title{Photon phonon entanglement in coupled optomechanical arrays}

\author{Uzma Akram}
\email{uzma@physics.uq.edu.au}
\affiliation{Centre for Engineered Quantum Systems, School of Mathematics and Physics, The University of Queensland, St Lucia, QLD 4072, Australia}
\author{William Munro}
\affiliation{NTT Basic Research Laboratories, NTT Corporation, 3-1 Morinosato-Wakamiya, Atsugi, Kanagawa 243-0198, Japan}
\affiliation{National Institute of Informatics, 2-1-2 Hitotsubashi, Chiyoda-ku, Tokyo-to 101-8430, Japan}
\author{Kae Nemoto}
\affiliation{National Institute of Informatics, 2-1-2 Hitotsubashi, Chiyoda-ku, Tokyo-to 101-8430, Japan}
\author{G. J. Milburn}
\affiliation{Centre for Engineered Quantum Systems, School of Mathematics and Physics, The University of Queensland, St Lucia, QLD 4072, Australia}

\begin{abstract}
We consider an array of three optomechanical cavities coupled either reversibly or irreversibly to each other and calculate the amount of entanglement between the different optical and mechanical modes. We show the composite system exhibits intercavity photon-phonon entanglement.
\end{abstract}

\maketitle
\section{Introduction}
Quantum cavity optomechanics capitalizes on the radiation pressure of light exerted on a mechanical degree of freedom  in a cavity  ~\cite{radpress1, radpress2, radpress3} to enable coupling between these different degrees of freedom. This currently highly active field of quantum physics is witnessing accelerated theoretical development~\cite{kippenbergreview, APS-Physics, isart1, isart2, isart3, chang, regal} as well as significant experimental achievements~\cite{corbitt, Schliesser, eichenfield, Thompson, markusstrongcoupling,vacoptomechrate}. More recently, an efficient quantum interface between optical photons and mechanical phonons has been illustrated ~\cite{verhagen}. Optomechanics serves as an excellent test bed for fundamental experiments at the quantum-classical boundary, leading to innovative ways of controlling the mutual interaction between light and the mechanical motion of mesoscopic objects. 
Sideband cooling to the ground state of mechanical resonator systems ~\cite{Marquardt, Wilson-Rae, Genes} is a challenge experimentally, due to the unavoidable thermal coupling of the resonators to their
environments. Nevertheless these challenges have been overcome, with different experimental
groups recently demonstrating cooling to the ground state of mechanical oscillators in
cavity electromechanical and optomechanical systems~\cite{Connell, Kippenberg2011, Teufel2011, painter2011}. Hence we are now in the era of quantum mechanical control of macroscopic objects~\cite{aspelmeyerreview}. 

With the complete control of a single optomechanical system in sight, these systems can provide new tools for implementing quantum measurement schemes~\cite{Sahar,Ludwig} and applications of a quantum photon-phonon interface~\cite{amir,stannigel2}.  Hence it is important to investigate the existence of entanglement between photons and phonons when we couple several of these systems together. Currently, the mechanical and optical modes within one optomechanical system have been shown to exhibit a considerable degree of entanglement~\cite{vitali, muller, hartmann, vitali2}. The entanglement between the output optical fields of a trapped-mirror-system has also been described~\cite{wipf}. A remarkable feature of optomechanical entanglement is that it can be present even at non zero temperatures.  Mazzola and Paternostro~\cite{paternostro}, showed the presence of entanglement between a pair of optomechanical cavities, in the linearised regime, that arise when each cavity is driven by one of the twin beams generated by a source of spontaneous parametric down conversion. The optomechanical systems become entangled due to quantum correlations in the light sources driving each cavity. Here we are going to take a different approach and examine various entanglement properties between three different optomechanical cavities coupled via their optical ports in either a reversible or irreversible (cascaded) manner.  An alternative approach would be to couple the mechanical resonators rather than the optical resonators~\cite{Heinrich}. A system of two coupled microwave cavities containing a mechanical element has also been considered recently by Heinrich and Marquardt~\cite{heinrichmarquardt}. The irreversible coupling has also recently been used to prepare entangled states in a driven cascaded quantum optical network,~\cite{stannigel}. We note a similar scheme for generating distant optomechanical entanglement using reversible coupling between two cavities has recently been reported,~\cite{joshi}. However this approach, unlike ours uses an 
adiabatic approximation resulting in quite different dynamics.

This paper is structured as follows. We begin with a description of the model in section~\ref{theory} describing the two couplings (reversible and irreversible) in which the optomechanical array can be set up. We then treat each coupling configuration in detail in Section~\ref{RRFFarrays} for different choices of the driving laser frequencies: Firstly in sections~\ref{RRfull} and ~\ref{MEapproach} we analyse the dynamics when each optomechanical cavity in the array is driven by a laser field of the same frequency. This detuning allows both squeezing and beam splitter interactions between the optomechanical modes within each cavity. Our results show how the entanglement generated between intracavity modes can be distributed over intercavity modes in the presence of optical coupling between the different optomechanical cavities. Secondly, for each coupling configuration we consider the case when the driving lasers of each cavity have different frequencies in sections~\ref{secRRbluered} and~\ref{bluered}.  We then choose particular detunings with respect to the mechanical frequency of each optomechanical unit to ensure that the source cavity is driven on the blue sideband and the receiver cavity is driven on the red sideband. This results in the field in the source cavity becoming entangled with the mechanical resonator in that cavity. The entanglement is then transferred to the mechanical resonator of the receiver cavity via the red sideband. We show that in this case the composite system exhibits steady state intercavity entanglement under the stability conditions for each optomechanical cavity. Finally we summarise our results in section~\ref{summary}. 

\section{The Model}
\label{theory}

The basic idea in quantum cavity optomechanics is to induce a reversible coupling between an optical and mechanical resonator. Typically the interaction arises from the radiation pressure of light. The usual set up is modelled as an optical cavity with its resonance frequency altered by the displacement of some mechanical resonator. A shift in the resonance frequency of the optical cavity changes the circulating power and thus changes the radiation pressure on the mechanical resonator,  yielding the optomechanical coupling which gives rise to a plethora of effects depending on how the various parameters and configurations are manipulated in the system. For example, the cavity can be driven to a steady state amplitude and the nonlinear optomechanical interaction linearised around this amplitude. This gives a coupling that is quadratic in the amplitude of the optical and mechanical resonator.  The goal in recent experiments has been to push the macroscopic mechanical elements of optomechanical systems towards the quantum limit by various passive and active cooling protocols.

\begin{figure}[!h]
\centering
\includegraphics[scale=0.7]{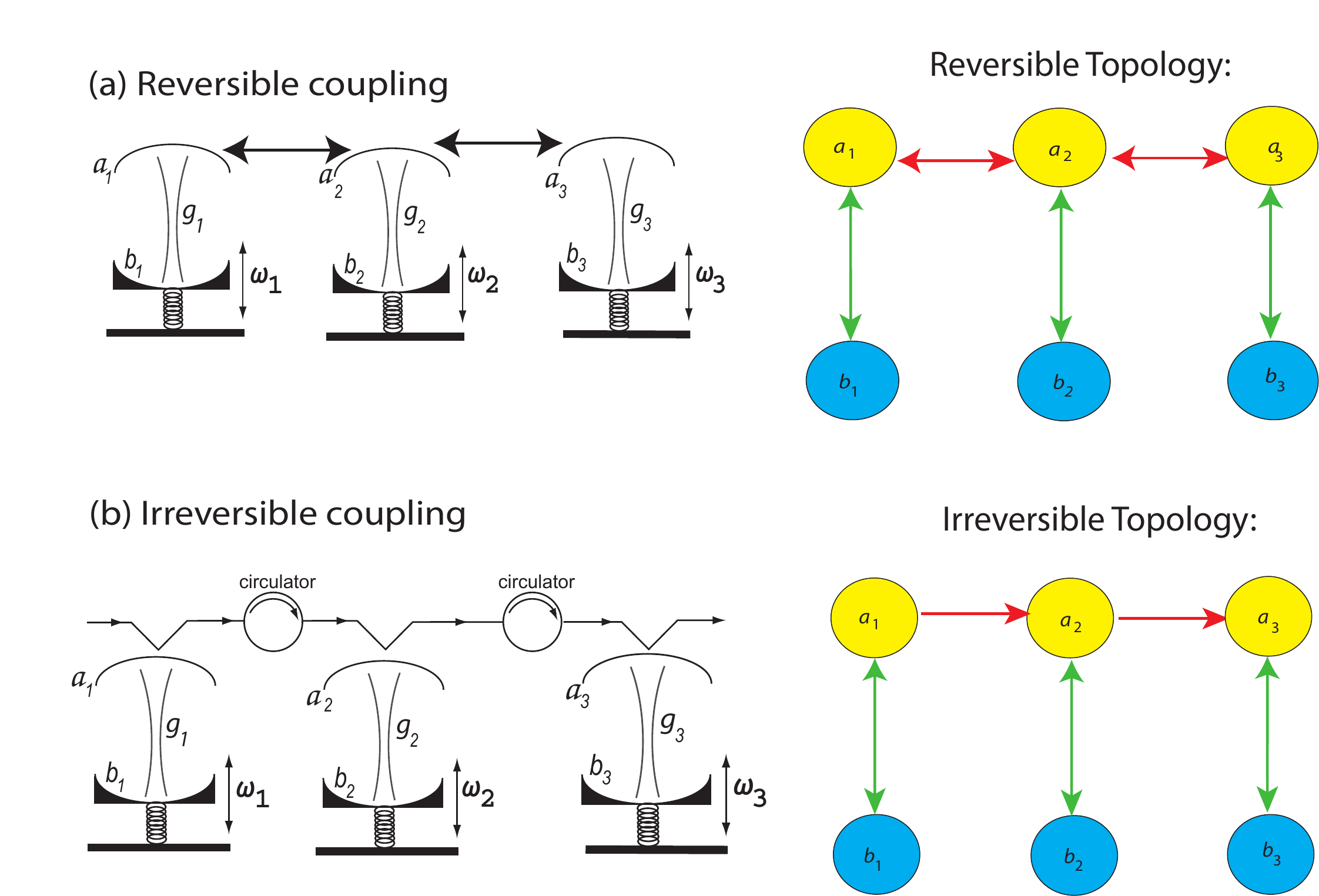}
\caption{(Color online) Model: The three optomechanical cavities can be coupled either (a)reversibly via coherent photon exchange (tunneling) between nearest neighbours or (b)irreversibly as forward feed using circulators. The array exhibits a different topology for each coupling configuration. }
\label{3cavitiesmodel}
\end{figure}

A system of three coupled cavities could be considered as either a triangular or a linear topology for the optomechanical array. We choose to consider a composite system of a linear chain of two or three identical cavities where the optical and mechanical modes in each cavity are quadratically coupled to each other. Each cavity ($1$,$2$,$3$) is an optical Fabry-Perot cavity in which one of the mirrors is subject to a harmonic restoring force and can thus move due to radiation pressure. The trio of optomechanical cavities can be coupled together in two different configurations: via a {\bf reversible} coupling or an {\bf irreversible} coupling configuration as illustrated in Fig.~\ref{3cavitiesmodel}. The mechanical resonator in both cases has a frequency $\omega_{m_{1,2,3}}$ and damping rate $\mu_{1,2,3}$ while the optical cavity has a resonance frequency $\omega_{c_{1,2,3}}$ and is strongly driven with a coherent pump field at frequency $\omega_{L_{1,2,3}}$. 

Reversible coupling allows a coherent photon exchange between nearest neighbour optical ports whereas the irreversible coupling is a forward feed method modelled using the cascaded systems approach~\cite{Car,gardiner}.  Reversible coupling requires that the cavities be evanescently coupled and thus must be close to each other, however the irreversible (cascaded) coupling does not require this, but will require circulators to be placed between the optical cavities to ensure irreversibility.

The total Hamiltonian for the uncoupled optomechanical array in the Schr\"{o}dinger picture is given by
\begin{eqnarray}
H & = & \hbar \sum_{k=1}^3 \omega_{c_{k}} a^{\dagger}_{k} a_{k}+\omega_{m_{k}} b^{\dagger}_{k} b_{k}+ G_{0}a^{\dagger}_{k} a_{k}(b_{k}+b^{\dagger}_{k}) +\left [E_{k} a_{k}^\dagger e^{-i\omega_{L_{k}} t}+E_{k}^* a_{k} e^{i\omega_{L_{k}} t}\right  ]
\label{totHam}
\end{eqnarray}
where $a_{k}(b_{k})$ and $a_{k}^\dagger(b_{k}^\dagger)$ are the annihilation and creation operators respectively of each optical (mechanical) mode and $E_{k}$ is the amplitude of the optical driving field of the {\em kth} cavity. 
We have assumed here that the cavities are identical such that the optomechanical coupling $G_{0}$ is the same. The master equation for the composite coupled system is given as,
\begin{eqnarray}
\frac{d\rho}{dt} & = &\frac{-i}{\hbar}[H,\rho]+\sum_{k=1}^3 \kappa_{k}{\cal D}[a_{k}]\rho +\mu_{k}(\bar{n}+1){\cal D}[b_{k}]\rho + \mu_{k}\bar{n}{\cal D}[b_{k}^\dagger]\rho \\\nonumber
& & +{\cal L}_{rev/irr}
\label{Meq} 
\end{eqnarray}
where the optical bath modes are taken to be in the vacuum state and $\bar{n}=1/(e^{\frac{\hbar\omega_{m}}{k_{B}T}}-1)$ is the mechanical phonon bath occupation number. The linewidth of cavity-$k$ is given by $\kappa_{k}$ and the mechanical damping rate is $\mu_{k}$ . The damping superoperator ${\cal D}[A]$ is defined by
\begin{equation}
{\cal D}[A]\rho=A\rho A^\dagger -\frac{1}{2} (A^\dagger A\rho+\rho A^\dagger A)
\end{equation} 
 ${\cal L}_{rev/irr}$ is the coupling between the optomechanical cavities which can either be ${\cal L}_{rev}$, reversible or ${\cal L}_{irr}$, irreversible. The reversible coupling is given as 
\begin{eqnarray}
{\cal L}_{rev}& = & -i\chi_{12}[a_{1}a_{2}^{\dagger}+a_{1}^\dagger a_{2},\rho] -i\chi_{23}[a_{2}a_{3}^{\dagger}+a_{3}^\dagger a_{2},\rho]
\label{rev}
\end{eqnarray}
where $\chi_{jk}$ is an arbitrary coupling strength between the reversibly coupled nearest neighbour cavities $j$ and $k$. The irreversible coupling is a feed forward exchange only between the optical modes described by the cascaded systems approach~\cite{Car,gardiner,gard_zoller} as
\begin{eqnarray}
{\cal L}_{irr}&= & \sqrt{\kappa_{1}\kappa_{2}}([a_{1}\rho,a_{2}^{\dagger}]+[a_{2},\rho a_{1}^{\dagger}]) +\sqrt{\kappa_{1}\kappa_{3}}([a_{1}\rho,a_{3}^{\dagger}]+[a_{3},\rho a_{1}^{\dagger}])  \\\nonumber
& & +\sqrt{\kappa_{2}\kappa_{3}}([a_{2}\rho,a_{3}^{\dagger}]+[a_{3},\rho a_{2}^{\dagger}]) 
\label{irr}
\end{eqnarray}
We assume that the intercavity coupling is either reversible or irreversible but not both together. As each cavity is externally driven, we linearise the radiation pressure interaction about the steady state field amplitude in each cavity. We now consider in detail both coupling configurations under the effect of different optomechanical interactions induced between the mechanical and optical modes of each cavity in the optomechanical array.   
 
 \section{Coupled optomechanical arrays}
 \label{RRFFarrays}
 
 \subsection{Reversible coupling}
 \label{RRcoupling}
 
We first consider the reversible coupling configuration between a chain of up to three optomechanical cavities. Here the three cavities are evanescently coupled to nearest neighbours with an arbitrary coupling strength $\chi_{jk}$ as given in Eq.~\ref{rev}. The topology we are considering here thus allows for a reversible exchange of optical excitations between the cavities as shown in Fig.~\ref{3cavitiesmodel}a. We are interested in the presence and possible distribution of entanglement between inter optical and mechanical degrees of freedom across the array.

Starting from Eq.~\ref{totHam}, we move to the interaction picture using the unitary transformation, 
\begin{equation}
U_0(t)=e^{-i\hbar(\omega_{L_1} a_1^\dagger a_1+\omega_{L_2} a_2^\dagger a_2 +\omega_{L_3} a_3^\dagger a_3)t}
\end{equation}
which will give rise to a detuning,  $\Delta_{k}=\omega_{c_{k}}-\omega_{L_{k}}$ between each cavity with respect to its corresponding driving laser frequency.  Hence the steady state cavity amplitudes, in the absence of the optomechanical interaction are given by
\begin{equation}
\alpha_k=\frac{-iE_k}{\kappa_k/2+i\Delta_k}
\end{equation}
Following the canonical transformation in the displaced reference frame, $\bar{a}_k=a_k-\alpha_k$, results in the effective Hamiltonian of the form 
\begin{eqnarray}
\label{HIrevers}
H_{I} &=&\hbar \sum_{k=1}^3 \Delta_{k} \bar{a}_{k}^{\dagger} \bar{a}_{k}+\omega_{m_k} b_{k}^\dagger b_k + g_k (\bar{a}_{k}^{\dagger} +\bar{a}_k) (b_{k}+b_{k}^{\dagger}) \\\nonumber
 &&+\hbar\chi_{12}\left ( \alpha_{1}^{*} \bar{a_{2}} e^{i(\omega_{L_1}-\omega_{L_2})t} +\alpha_{1} \bar{a}_{2}^{\dagger} e^{-i(\omega_{L_1}-\omega_{L_2})t} \right ) \\\nonumber 
&&+ \hbar\chi_{23}\left ( \alpha_{2}^{*} \bar{a_{3}} e^{i(\omega_{L_2}-\omega_{L_3})t} +\alpha_{2}\bar{a}_{3}^{\dagger} e^{-i(\omega_{L_2}-\omega_{L_3})t} \right) 
\end{eqnarray}
where $g_{k}=\alpha_{k}G_{0}$ is the effective optomechanical coupling strength proportional to the steady state amplitude of the cavity field due to linearisation of the radiation pressure force. We note an additional driving term on each optical mode due to the steady state coherent field leaking from its nearest neighbour. This can be filtered out using a beam splitter and a coherent local oscillator, and we thus ignore it in our work. 
The resulting linearised master equation in the interaction picture for the system of three reversibly coupled optomechanical cavities is now,
\begin{eqnarray}
\frac{d\rho}{dt}&=& -\frac{i}{\hbar}[H_I,\rho]+\sum_{k=1}^3\kappa_{k}{\cal D}[\bar{a}_k]\rho  +\mu_{k}(\bar{n}+1){\cal D}[b_{k}]\rho+\mu_{k}\bar{n}{\cal D}[b_{k}^\dagger]\rho \\ \nonumber 
&&+\chi_{12}\left ([\bar{a}_{1}\rho,\bar{a}^{\dagger}_{2}] e^{-i(\omega_{L_1}-\omega_{L_2})t}+[\bar{a}_{2},\rho \bar{a}_{1}^{\dagger}]e^{i(\omega_{L_1}-\omega_{L_2})t}\right )\\ \nonumber 
&&+\chi_{23}\left ([\bar{a}_{2}\rho,\bar{a}_{3}^{\dagger}] e^{-i(\omega_{L_2}-\omega_{L_3})t}+[\bar{a}_{3},\rho \bar{a}_{2}^{\dagger} ]e^{i(\omega_{L_2}-\omega_{L_3})t}\right ) 
\label{MERR}
\end{eqnarray}
where the reversible coupling terms introduced in Eq.~\ref{rev} have been considered. The reversible coupling is analogous to a beam splitter interaction between two modes. We have ignored any phase differences between the relative driving fields here.  We will drop the bars from now on, however we are working in the displaced picture as evidenced by the appearance of the effective coupling strength $g_{k}$ in the interaction Hamiltonian, $H_{I}$. In the following we consider  different choices for the driving laser frequencies, such that each cavity may be blue or red detuned.

\subsubsection{Equal driving laser frequencies, with the full interaction Hamiltonian}
\label{RRfull}
We now consider the dynamics of the chain of reversibly coupled optomechanical systems such that the composite system evolves as given by the full interaction Hamiltonian $H_{I}$, in Eq.~\ref{HIrevers}. We choose the frequencies on all driving laser fields to be the same $\omega_{L_1}=\omega_{L_2} =\omega_{L_3}=\omega_{L}$. We can tune all driving fields to be simultaneously on the same sideband (i.e. all red or all blue). The  master equation  for this choice of equal laser frequencies is now,
\begin{eqnarray}
\frac{d\rho}{dt}&=& -\frac{i}{\hbar}[H_I,\rho]+\sum_{k=1}^3\kappa_{k}{\cal D}[\bar{a}_k]\rho  +\mu_{k}(\bar{n}+1){\cal D}[b_{k}]\rho+\mu_{k}\bar{n}{\cal D}[b_{k}^\dagger]\rho \\ \nonumber 
&&+\chi_{12}\left ([\bar{a}_{1}\rho,\bar{a}^{\dagger}_{2}]+[\bar{a}_{2},\rho \bar{a}_{1}^{\dagger}]\right ) +\chi_{23}\left ([\bar{a}_{2}\rho,\bar{a}_{3}^{\dagger}] +[\bar{a}_{3},\rho \bar{a}_{2}^{\dagger} ]\right ) 
\label{MERRnew}
\end{eqnarray}
such that there are no time dependent coeffcients accompanying the reversible coupling terms. Note that this choice of laser frequencies requires that either $\omega_{L_{k}}=\omega_{c_{k}}-\omega_{m_{k}}$ (all cavities on the red sideband, $\Delta_{k}=\omega_{m_{k}}$) or $\omega_{L_{k}}=\omega_{c_{k}}+\omega_{m_{k}}$ (all cavities on the blue sideband, $\Delta_{k}=-\omega_{m_{k}}$).

As has been shown previously~\cite{vitali}, intracavity photon-phonon entanglement is present within each optomechanical cavity. However here we are interested  in the presence of entanglement between the optical and mechanical resonators in distinct optomechanical cavities. Our composite system under study is in a Gaussian state as it starts from the vacuum, and the equations of motion are linear. Consequently, to quantify the entanglement, we can use the logarithmic negativity measure for Gaussian states formulated by Vidal~\cite{vidal}. The logarithmic negativity between two states is expressed in terms of the entries of their covariance matrix, $\gamma$ which is a $4 \times 4$ matrix given as 
\begin{equation}
\gamma=\left(
\begin{matrix}
\gamma_{A} & \gamma_{C} \\
\gamma_{C}^{T} & \gamma_{B}  \\
\end{matrix} \right), \quad \gamma_{A},\gamma_{B}, \gamma_{C} \in M(2,\Re)
\end{equation}
The matrices $\gamma_{A,B}$ arise from position and momentum quadratures of the optical and mechanical modes respectively, while $\gamma_{C}$ is a result of cross terms between the optical and mechanical position and momentum quadratures. The logarithmic negativity is then calculated as 
\begin{equation}
E_{N}=
\begin{cases}
-\log f(\gamma)/2, & \text{if}\; f(\gamma)<1,\\
0,& \text{otherwise.} \\
\end{cases}
\label{logneg}
\end{equation}

where the function, $f:C(4)\longrightarrow \Re^{+}$ is defined as 
\begin{equation}
f(\gamma)=\Gamma_{A,B,C}-\sqrt{\Gamma_{A,B,C}^{2}-|\gamma|}
\end{equation}
such that 
\begin{equation}
\Gamma_{A,B,C}=\frac{\mid\gamma_{A}\mid+\mid\gamma_{B}\mid}{2}-\mid\gamma_{C}\mid
\end{equation}
where $|\gamma|$ is the determinant of $\gamma$. 

Hence in order to determine the matrix elements of each $2\times 2$ matrix, $\gamma_{A,B,C}$, we calculate the second order moments from the master equation of the composite system. Inserting the relevant second order moments into Eq.~\ref{logneg} we can determine the entanglement between any two modes of the system.
 \begin{figure}
 \centering
\includegraphics[scale=0.8]{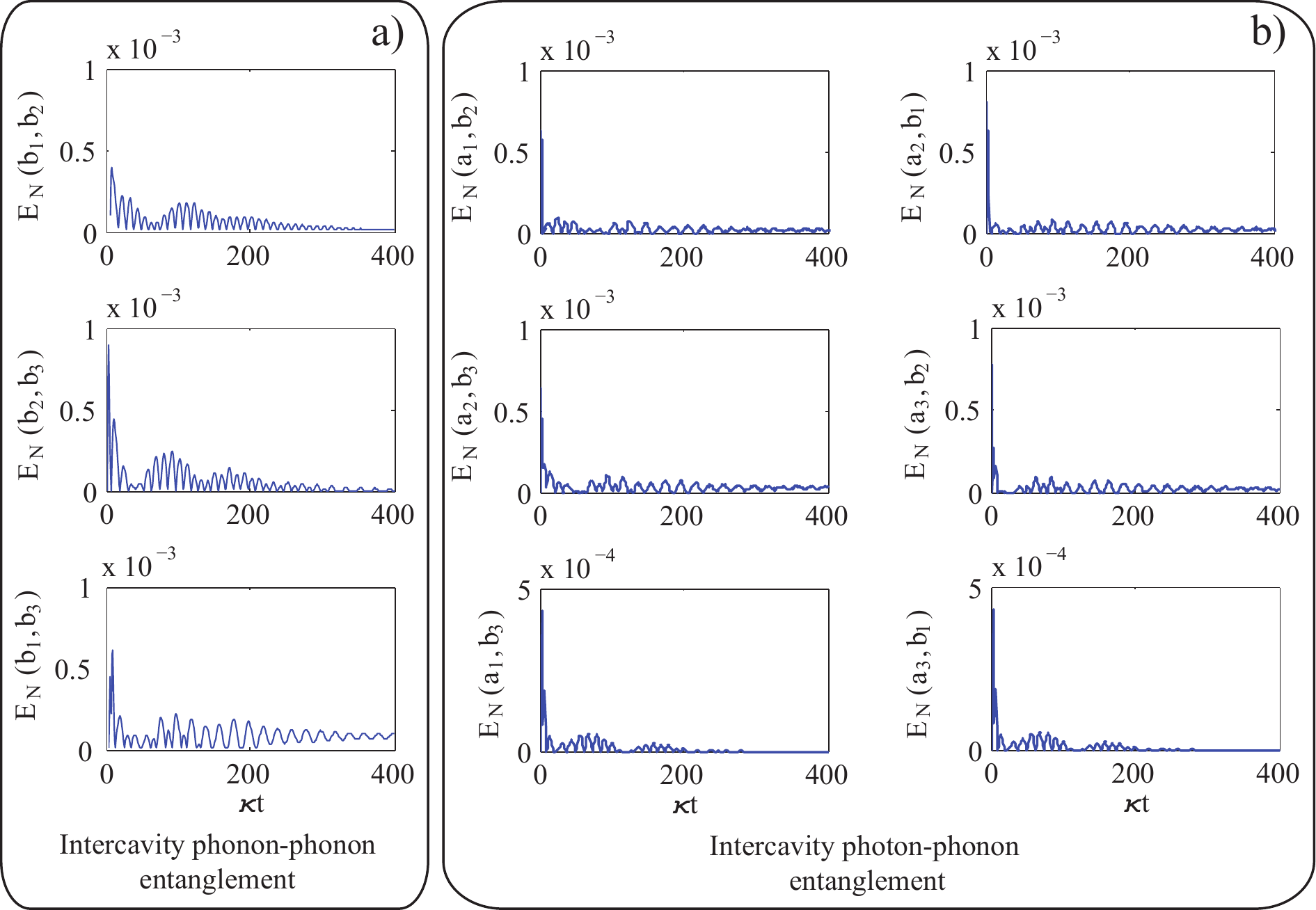}
\caption{{\em Reversible coupling with the full interaction} \\The temporal evolution of the entanglement, $E_{N}$ quantified by the logarithmic negativity  between all possible selections of intercavity modes of the optomechanical array(a) phonon-phonon and (b) photon-phonon where have chosen $\kappa_{1,2,3} = \kappa$, $\mu_{1,2,3}/\kappa=0.01, \Delta_{1,2,3}/\kappa=\omega_{1,2,3}/\kappa=200, g_{1,2,3}/\kappa=0.5,\chi_{12}/\kappa=\chi_{23}/\kappa=1$ and $\bar{n}=0 $.}
  \label{3RRasbs}
\end{figure}

Fig.~\ref{3RRasbs}(a) shows the entanglement quantified by the logarithmic negativity between the intercavity mechanical modes as a function of time where we have all parameters in units of the cavity linewidth, $\kappa$. We see that the mirrors of different optomechanical cavities in the array are entangled in the absence of a direct coupling induced between them. Fig.~\ref{3RRasbs}(b) shows the entanglement quantified by the logarithmic negativity between all two mode intercavity photon phonon combinations of the composite system of coupled optomechanical cavities.  Clearly all the six pairs of photon phonon modes of the system are entangled with each other. We note the symmetry for pairs of oscillators $a_{j},b_{k}$ and $a_{k},b_{j}$. This is due to the symmetry in the Hamiltonian, Eq.~\ref{HIrevers} for cavities with identical parameters $\omega_{m_{k}}$ and $g_{k}$. 

\begin{figure}[!h]
\centering
\includegraphics[scale=0.7]{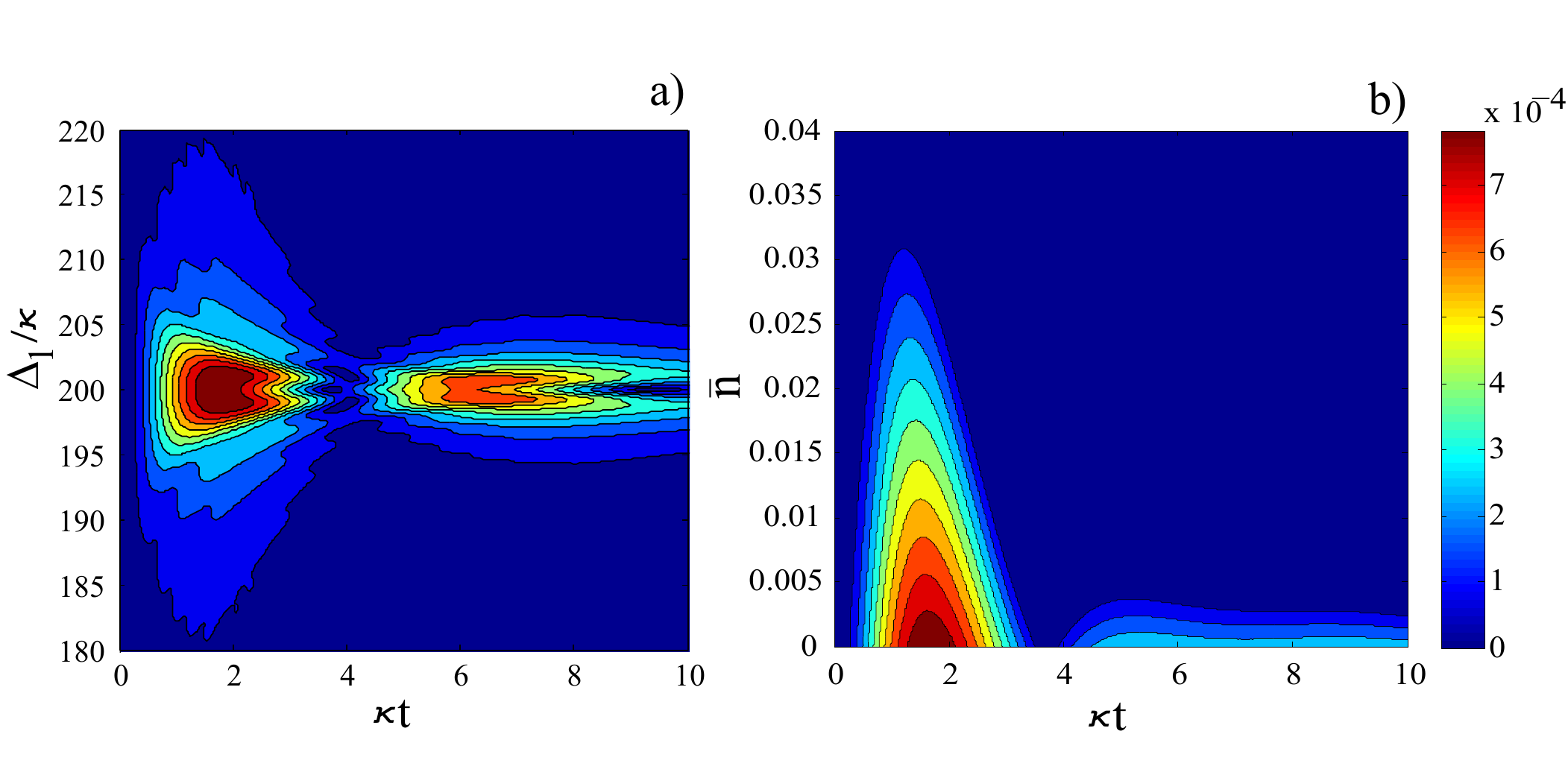}
\caption{(Color online) {\em Reversible coupling with the full interaction} \\The temporal evolution of the entanglement, $E_{N}$ quantified by the logarithmic negativity  between intercavity phonon modes $b_{1},b_{2}$ of the optomechanical array as a function of (a) detuning of cavity 1, $\Delta_{1}$, and (b) thermal occupation number $\bar{n}$ where we have chosen $\kappa_{1,2,3} = \kappa$, $\mu_{1,2,3}/\kappa=0.01$, $\omega_{1}/\kappa=200$, $\Delta_{2,3}/\kappa=\omega_{2,3}/\kappa=200$, $g_{1,2,3}/\kappa=0.5$, $\chi_{12}/\kappa=\chi_{23}/\kappa=1$.} 
\label{deldthermalRR}
\end{figure} 
When the three cavities are each detuned to the red sideband, we have found the entanglement reaches a maximum (as a function of time) when the detuning is resonant with the mechanical frequency of each cavity, i.e. $\Delta_{k}=\omega_{m_{k}}$ as illustrated in Fig.~\ref{deldthermalRR}(a) for $\Delta_{1}$. We would like to remind the reader, that even though this is red sideband detuning, the full linearised interaction Hamiltonian has been considered here, (that is to say we do not make the rotating wave approximation (RWA) at this point) so that the squeezing terms $\hat{a}_{k}\hat{b}_{k}$ and $\hat{a}_{k}^{\dagger}\hat{b}_{k}^{\dagger}$ also contribute towards the dynamics of the system and in fact are responsible for the presence of the entanglement within each optomechanical unit. Detuning the cavity to the blue sideband, $\Delta=-\omega_m$, one would expect to achieve strong entanglement. However, for the blue sideband driving, care should be taken to ensure that the steady state, about which we have linearised the radiation pressure interaction, remains stable.  If we operate in a regime in which the RWA would be valid, the steady state on the blue sideband is unstable when  $g < \sqrt{\kappa\mu}/2$ ~\cite{optomechentanglement_vitali}.  Moreover because we are operating in the resolved sideband regime, $\omega_m >> \kappa, \mu$, the strength of the optomechanical coupling in each cavity, $g_{k}$, is restricted to very small values. As a result, for such weak coupling parameters, the intracavity entanglement does not transfer to intercavity modes on the blue sideband. 

If we tune to the red sideband, $\Delta=\omega_{m}$, so that the beam splitter interaction is resonant, one would not expect much optical-mechanical entanglement within the RWA. However for red sideband detuning the stability conditions give  $g<\frac{1}{2}\sqrt{\omega_m^2+\frac{\mu^2 +\kappa^{2}}{4}}$. If the mechanical frequency, $\omega_{m}$, is large enough this enables one to use large values of $g$ and remain within a stable operating regime. Under those conditions it would not be valid to make the RWA and thus the non resonant terms cannot be neglected resulting in entanglement between the optical and mechanical modes even though we are driving on the red sideband. This results in an exchange of excitations on the optomechanical branches of the chain of oscillators, consequently distributing the intracavity entanglement between oscillators of different cavities due to the coupling between the optical ports. 
Hence the coupling between the optomechanical units facilitates the distribution of intracavity entanglement over the intercavity optical and mechanical modes.

Fig.~\ref{deldthermalRR}(b) shows how the intercavity phonon entanglement varies with an increase in thermal phonon number, $\bar{n}$ of the mechanics. Clearly each mechanical mode of the coupled optomechanical array will need to be as close as possible to its ground state to maintain the intercavity phonon phonon entanglement induced by the reversible coupling between the optical modes.

\subsubsection{Unequal driving laser frequencies}
\label{secRRbluered}
 So far we have looked at the distribution of entanglement in optomechanical arrays in the reversible coupling under the full interaction Hamiltonian, such that optomechanical entanglement was generated within each cavity independently.  We are now interested in the distribution of entanglement in optomechanical arrays given we have entanglement generated only in the source cavity by tuning to the blue sideband. For simplicity we consider a chain of only two coupled cavities here, however our results can be extended for a large number of similarly coupled optomechanical systems.  We choose the source cavity to be exclusively on the blue sideband and the receiver cavity to be only on the red sideband. 

We  consider the cavities to be on resonance so that $\omega_{c_1}=\omega_{c_{2}}=\omega_{c}$, and assume the mechanical resonators have the same frequency, $\omega_{m_1}=\omega_{m_2}=\omega_{m}$, but as we drive cavity 1 on the blue sideband, $\Delta_{1}=\omega_{c}-\omega_{L_{1}}=-\omega_{m}$, and cavity 2 on the red sideband, $\Delta_{2}=\omega_{c}-\omega_{L_{2}}=\omega_{m}$. This choice then implies, $\omega_{c}=\frac{\omega_{L_1}+\omega_{L_2}}{2}$ for the optical frequencies and $\omega_{m}=\frac{(\omega_{L_1}-\omega_{L_2})}{2}$ for mechanical frequencies. However there will be a detuning created by the different driving laser frequencies, $\omega_{L_1}\neq\omega_{L_2}$. We now go to another interaction picture with respect to the mechanical frequency, $\omega_{m}$ and make the RWA so that the 
interaction Hamiltonian of the composite system for this coupling configuration is 
\begin{eqnarray}
\acute{H}_{I}&=&\hbar g_{1}(a_{1}b_{1} + a_{1}^{\dagger}b_{1}^{\dagger}) + \hbar g_{2}(a_{2}^{\dagger}b_{2}+a_{2}b_{2}^{\dagger})
\label{HIprime}
\end{eqnarray}
and the master equation in the RWA now will be given as
\begin{eqnarray}
\frac{d\rho}{dt}&=& -\frac{i}{\hbar}[\acute{H}_I,\rho]+\sum_{k=1}^2\kappa_k{\cal D}[a_k]\rho +\mu_{k}(\bar{n}+1){\cal D}[b_{k}]\rho+\mu_{k}\bar{n}{\cal D}[b_{k}^\dagger]\rho \\ \nonumber
&&  -i\chi_{12}\left[a_1 a_2^\dagger +a_2 a_1^\dagger, \rho\right]
\label{blueredME_RR}
\end{eqnarray}
Here in the source cavity, the optomechanical interaction is a two-mode squeezing interaction
which will entangle the mechanics and the field. Tuning the receiver optomechanical cavity to the red sideband, we generate a beam splitter interaction which can swap the state of the mechanics and the field. Thus we can use the first cavity to entangle the mirror motion and the field, take the field out via the reversible coupling and then swap it into the mirror motion in the second cavity. Hence this system is capable of generating entanglement between the mechanical resonators, $b_{1}$ and $b_{2}$ even though the receiver cavity is explicitly on the red sideband. 

\begin{figure}[!h]
\centering
\includegraphics[scale=0.7]{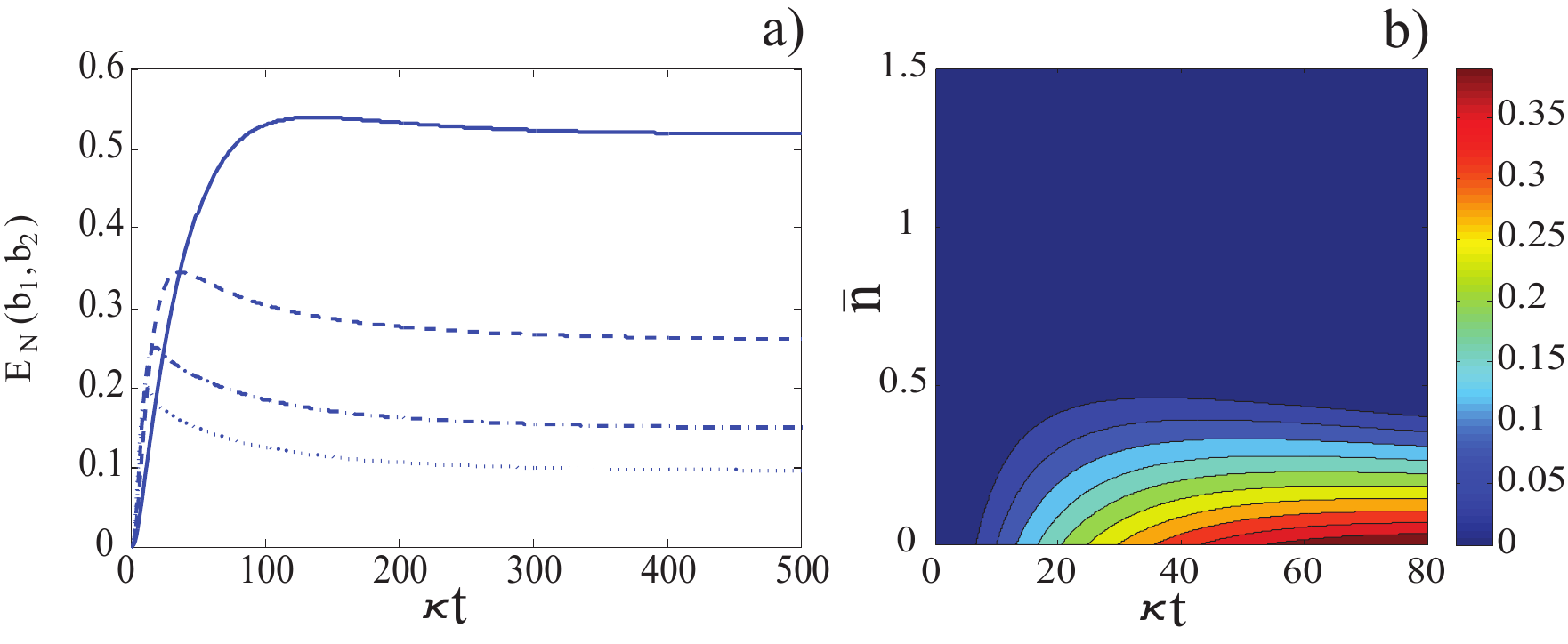}
\caption{(Color online) {\em Reversible coupling with cavity 1 blue detuned and cavity 2 red detuned.}\\ The temporal evolution of entanglement, $ E_{N}$ quantified by the logarithmic negativity  between the intercavity phonons of two optomechanical cavities ($b_{1}$,$b_{2}$) where we have chosen $\kappa_{1,2} = \kappa$, $\mu_{1,2}/\kappa=0.01$, $g_{1}/\kappa=0.02$, $\chi_{12}/\kappa=1$ (a) with $\bar{n}=0$ and different $g_{2}/\kappa$: 0.1 (solid line), 0.2 (dashed line), 0.3 (dashed dotted line) and 0.4 (dotted line) and (b) as a function of  thermal phonon number $\bar{n}$ with $g_{2}/\kappa=0.1$.}
\label{RRbluered}
\end{figure} 
We illustrate this in Fig.~\ref{RRbluered}(a), which shows how the entanglement varies between the two resonators $b_{1}$ and $b_{2}$ as a function of time with dimensionless parameters for different optomechanical coupling strengths $g_{2}$  for a fixed $g_{1}$. We note for this coupling configuration that the entanglement between intercavity phonons becomes larger for weaker optomechanical coupling of cavity 2, which is on the red sideband. Fig.~\ref{RRbluered}(b) shows how the temporal intercavity phonon entanglement varies with the phonon occupation number $\bar{n}$. While there is still significant entanglement in this coupling for non zero $\bar{n}$, here again we find that the mechanical resonators of the coupled optomechanical system would need to remain close to their respective ground states to stay entangled with each other.  

\begin{figure}[!h]
\centering
\includegraphics[scale=0.8]{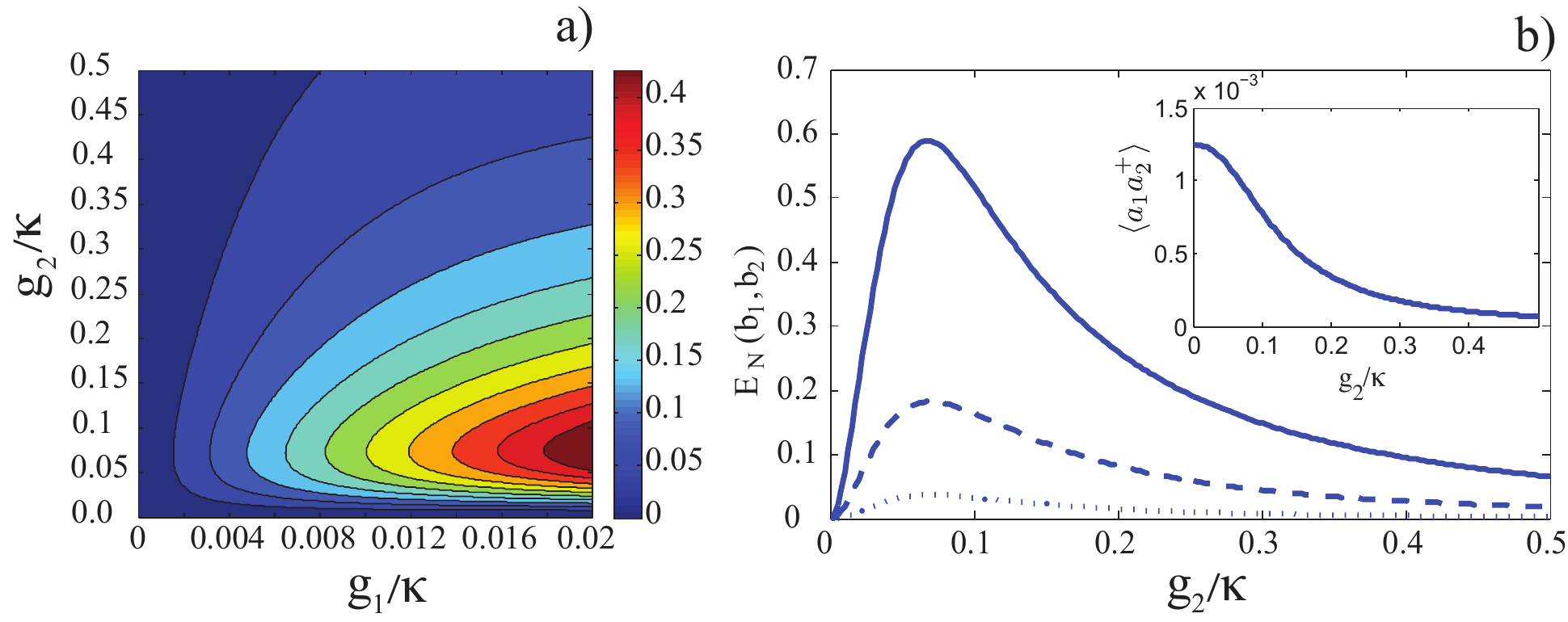}
\caption{(Color online) {\em Reversible coupling with cavity 1 blue detuned and cavity 2 red detuned.}\\ The steady state entanglement, $E_{N}$ quantified by the logarithmic negativity  between the intercavity phonons of two optomechanical cavities ($b_{1}$,$b_{2}$) where we have chosen $\kappa_{1,2} = \kappa$, $\mu_{1,2}/\kappa=0.01$ and $\bar{n}=0$, (a) as a function of $g_{1}/\kappa$ and $g_{2}/\kappa$ for $\chi_{12}/\kappa=1$ and (b) as a function of $g_{2}/\kappa$ for $g_{1}=0.02$ and different $\chi_{12}/\kappa$: 1 (solid line), 0.5 (dashed line) and 0.2 (dotted line). The inset shows the correlation $<a_{1}a^{\dagger}_{2}>$ as a function of $g_{2}/\kappa$ for the same parameters except for $\chi_{12}/\kappa=1$ and $g_{1}/\kappa=0.02$.}
\label{SSRRbluered}
\end{figure} 

Clearly from Fig.~\ref{RRbluered}(a), the entanglement in the composite system here is much larger in magnitude from the previous case of the full linearised interaction Hamiltonian. For this coupling configuration we find the system goes to a steady state and hence evaluate the steady state second order moments and calculate the logarithmic negativity between the mechanical modes for both cavities.  We plot the steady state entanglement quantified by the logarithmic negativity between the resonators $b_{1},b_{2}$ versus the two optomechanical coupling strengths, $g_{1}$ and $g_{2}$ in Fig.~\ref{SSRRbluered}(a). We choose parameters for both cavities such that the respective stability conditions are satisfied in the steady state. 

Finally for the reversible coupling configuration, Fig.~\ref{SSRRbluered}(b) shows the steady state entanglement between intercavity phonon modes increases with the strength of the reversible coupling $\chi_{12}$, between the optical modes. We  note from Fig.~\ref{SSRRbluered}(b) that the entanglement does not linearly increase with $g_{2}$. The correlation $<b_{2}b_{1}>$ and thus the intercavity entanglement arises from the state swap (arising from the form of the red sideband coupling) between $a_{2}$ and $b_{2}$, once the field has been transferred from the optical mode $a_{1}$ to $a_{2}$.  Initially as $g_{2}$ increases, the entanglement $EN(b_{1},b_{2})$ also increases. However, for larger $g_{2}$ the rate of the state swap between $a_{2}$ and $b_{2}$ becomes close to and eventually faster than the coupling between $a_{1}$ and $a_{2}$ which is restricted by the fixed coupling parameter, $\chi_{12}$. Consequently this results in a decay of the correlation $<a_{1}a^{\dagger}_{2}>$ and hence $EN(b_{1},b_{2})$ along the $g_{2}$ axis. As the results show, there is therefore an optimum combination of $g_{1}$ and $g_{2}$ for which the intercavity phonons are maximally entangled in the steady state before logarithmic negativity begins to decrease with increasing $g_{2}$.

\subsection{Forward feed (irreversible) coupling}
\label{ffcoupling}
We now investigate the presence and possible distribution of entanglement between intercavity modes in the forward feed coupling configuration.  Here the cavities are coupled via a unidirectional coupling only. In our calculations, we have neglected time delay between the cavities. Effectively, we can consider cavity $1$ as the "main source" cavity while cavity $3$ is only a receiver cavity. Cavity $2$ on the other hand receives photons from cavity $1$ as well as drives cavity $3$. The topology of our set up in this coupling configuration is further illustrated in Fig.~\ref{3cavitiesmodel}(b). Hence we have reversible interactions between the optomechanical branches of the chain but irreversible interactions between the optical ports coupling the individual optomechanical units. This is quite a different configuration to previous work on coupled oscillator arrays~\cite{eisertplenio}.
We have for the interaction Hamiltonian,
\begin{eqnarray}
\label{HI1}
H_{I} &=&\hbar \sum_{k=1}^3 \Delta_{k} \bar{a}_{k}^{\dagger} \bar{a}_{k}+\omega_{m_k} b_{k}^\dagger b_k + g_k (\bar{a}_{k}^{\dagger} +\bar{a}_k) (b_{k}+b_{k}^{\dagger}) \\ \nonumber
 &&+\hbar\sqrt{\kappa_{1}\kappa_{2}}\left ( \alpha_{1}^{*} \bar{a_{2}} e^{i(\omega_{L_1}-\omega_{L_2})t} +\alpha_{1} \bar{a}_{2}^{\dagger} e^{-i(\omega_{L_1}-\omega_{L_2})t} \right ) \\ \nonumber 
&&+ \hbar\sqrt{\kappa_{2}\kappa_{3}}\left ( \alpha_{2}^{*} \bar{a_{3}} e^{i(\omega_{L_2}-\omega_{L_3})t} +\alpha_{2}\bar{a}_{3}^{\dagger} e^{-i(\omega_{L_2}-\omega_{L_3})t} \right ) \\ \nonumber 
&&+\hbar\sqrt{\kappa_{1}\kappa_{3}}\left ( \alpha_{1}^{*} \bar{a}_{3} e^{i(\omega_{L_1}-\omega_{L_3})t} +\alpha_{1}\bar{ a}_{3}^{\dagger} e^{-i(\omega_{L_1}-\omega_{L_3})t} \right)
\end{eqnarray}
 where $\Delta_{k}=\omega_{c_{k}}-\omega_{L_{k}}$ is the detuning of each cavity with respect to the driving laser field. As before, we have an additional driving (last three terms in Eq.~\ref{HI1}) on each receiver cavity ($2,3$) due to the steady state coherent field leaking from each source cavity ($1,2$) which can be filtered out and we thus ignore it in our work. In Eq.~\ref{HI1}, $g_{k}$ is now the effective optomechanical coupling strength proportional to the steady state amplitude of the cavity field due to linearisation of the radiation pressure force. Following the canonical transformation in the displaced reference frame, $\bar{a}_k=a_k-\alpha_k$, the master equation in the interaction picture for the cascaded system of three optomechanical cavities is now,
\begin{eqnarray}
\frac{d\rho}{dt}&=& -\frac{i}{\hbar}[H_I,\rho]+\sum_{k=1}^3\kappa_k{\cal D}[\bar{a}_k]\rho  +\mu_{k}(\bar{n}+1){\cal D}[b_{k}]\rho+\mu_{k}\bar{n}{\cal D}[b_{k}^\dagger]\rho \\ \nonumber 
&& +\sqrt{\kappa_{1}\kappa_{2}}\left ([\bar{a}_{1}\rho,\bar{a}^{\dagger}_{2}] e^{-i(\omega_{L_1}-\omega_{L_2})t}+[\bar{a}_{2},\rho \bar{a}_{1}^{\dagger} ]e^{i(\omega_{L_1}-\omega_{L_2})t}\right )\\ \nonumber 
&&+\sqrt{\kappa_{2}\kappa_{3}}\left ([\bar{a}_{2}\rho,\bar{a}_{3}^{\dagger}] e^{-i(\omega_{L_2}-\omega_{L_3})t}+[\bar{a}_{3},\rho \bar{a}_{2}^{\dagger} ]e^{i(\omega_{L_2}-\omega_{L_3})t}\right ) \\ \nonumber 
&&+\sqrt{\kappa_{1}\kappa_{3}}\left ([\bar{a}_{1}\rho,\bar{a}_{3}^{\dagger}] e^{-i(\omega_{L_1}-\omega_{L_3})t}+[\bar{a}_3,\rho \bar{a}_1^\dagger ]e^{i(\omega_{L_1}-\omega_{L_3})t}\right )
\label{newME1}
\end{eqnarray}
 We have ignored any phase differences between the relative driving fields here.  From now on we drop the bars from the operators but we will be working in the displaced picture and hence consider the effective optomechanical coupling strength $g_{k}$. In the following we consider  different choices for the driving laser frequencies, such that each cavity may be blue or red detuned.  

\subsubsection{Equal driving laser frequencies with the full interaction in Hamiltonian}
\label{MEapproach}

Firstly, we will consider the three optomechanical systems coupled in a cascaded fashion such that the composite system evolves as given by the full interaction Hamiltonian $H_{I}$, in Eq.~\ref{HI1}. We choose the frequencies on all driving laser fields to be the same $\omega_{L_1}=\omega_{L_2} =\omega_{L_3}=\omega_{L}$. We can tune all driving fields to be simultaneously on the same sideband (i.e. all red or all blue). The master equation for this choice of equal laser frequencies is 
\begin{eqnarray}
\frac{d\rho}{dt}&=& -\frac{i}{\hbar}[H_I,\rho]+\sum_{k=1}^3\kappa_k{\cal D}[a_k]\rho +\mu_{k}(\bar{n}+1){\cal D}[b_{k}]\rho+\mu_{k}\bar{n}{\cal D}[b_{k}^\dagger]\rho \\ \nonumber
&&  + \sqrt{\kappa_1\kappa_2}\left ([a_1\rho,a_2^\dagger] +[a_2,\rho a_1^\dagger ]\right )\\\nonumber
&&+\sqrt{\kappa_2\kappa_3}\left ([a_2\rho, a_{3}^{\dagger} +[a_3,\rho a_2^\dagger ]\right) +\sqrt{\kappa_1\kappa_3}\left ([a_1\rho, a_3^\dagger] +[a_3,\rho a_1^\dagger ]\right)
\label{redME}
\end{eqnarray}
such that there are no time dependent coefficients accompanying the forward feed coupling terms. As in the reversible coupling case this choice of laser frequencies places the restriction that either $\omega_{L}=\omega_{c_k}-\omega_{m_k}$  (all red, $\Delta_{k}=\omega_{m_{k}}$) or $\omega_{L}=\omega_{c_k}+\omega_{m_k}$  (all cavities on the blue sideband, $\Delta_{k}=-\omega_{m_{k}}$). 

\begin{figure}[!h]
\centering
\includegraphics[scale=0.7]{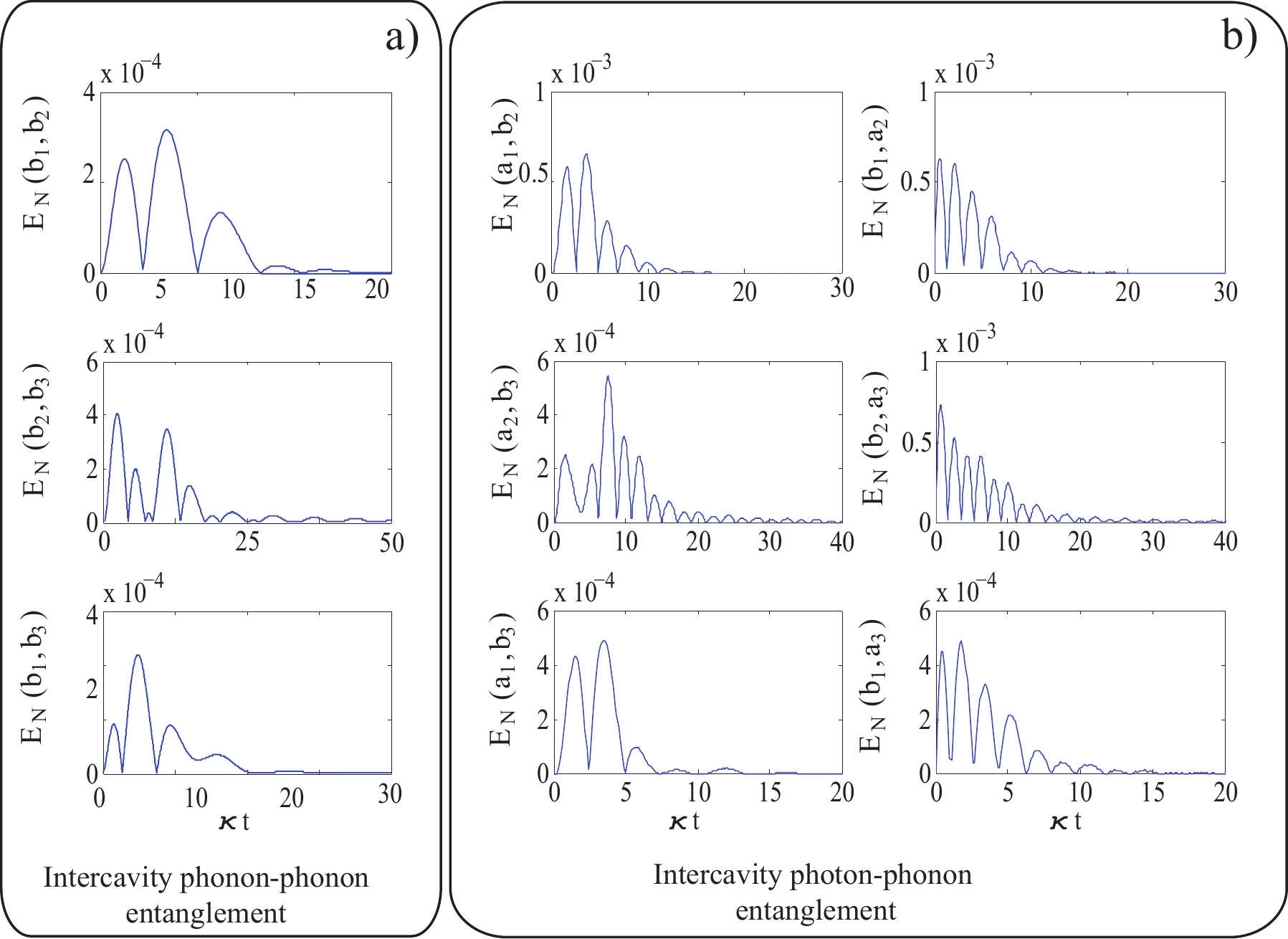}
\caption{{\em Forward feed coupling with the full interaction.}\\The temporal evolution of the entanglement, $E_{N}$ quantified by the logarithmic negativity between all possible intercavity modes of the optomechanical array (a) phonon-phonon and (b) photon-phonon where we have chosen $\kappa_{1,2,3}=\kappa$, $\mu_{1,2,3}/\kappa=0.01, \Delta_{1,2,3}/\kappa=\omega_{1,2,3}/\kappa=400, g_{1,2,3}/\kappa=0.5$ and $\bar{n}=0 $.}
\label{3bs_ff1}
\end{figure}

Fig.~\ref{3bs_ff1}(a) shows the existence of pair-wise entanglement between the intercavity mechanical modes $b_1$,$b_2$ and $b_3$ of the composite system for the forward feed coupling between the optical ports. 
Fig.~\ref{3bs_ff1}(b) shows the existence of entanglement between all possible pairs of the optical and mechanical modes of different cavities, i.e. intercavity photon phonon entanglement in the composite system.  Again all the six pairs of photon phonon modes of the system are entangled with each other. 

\begin{figure}[!h]
\centering
\includegraphics[scale=0.6]{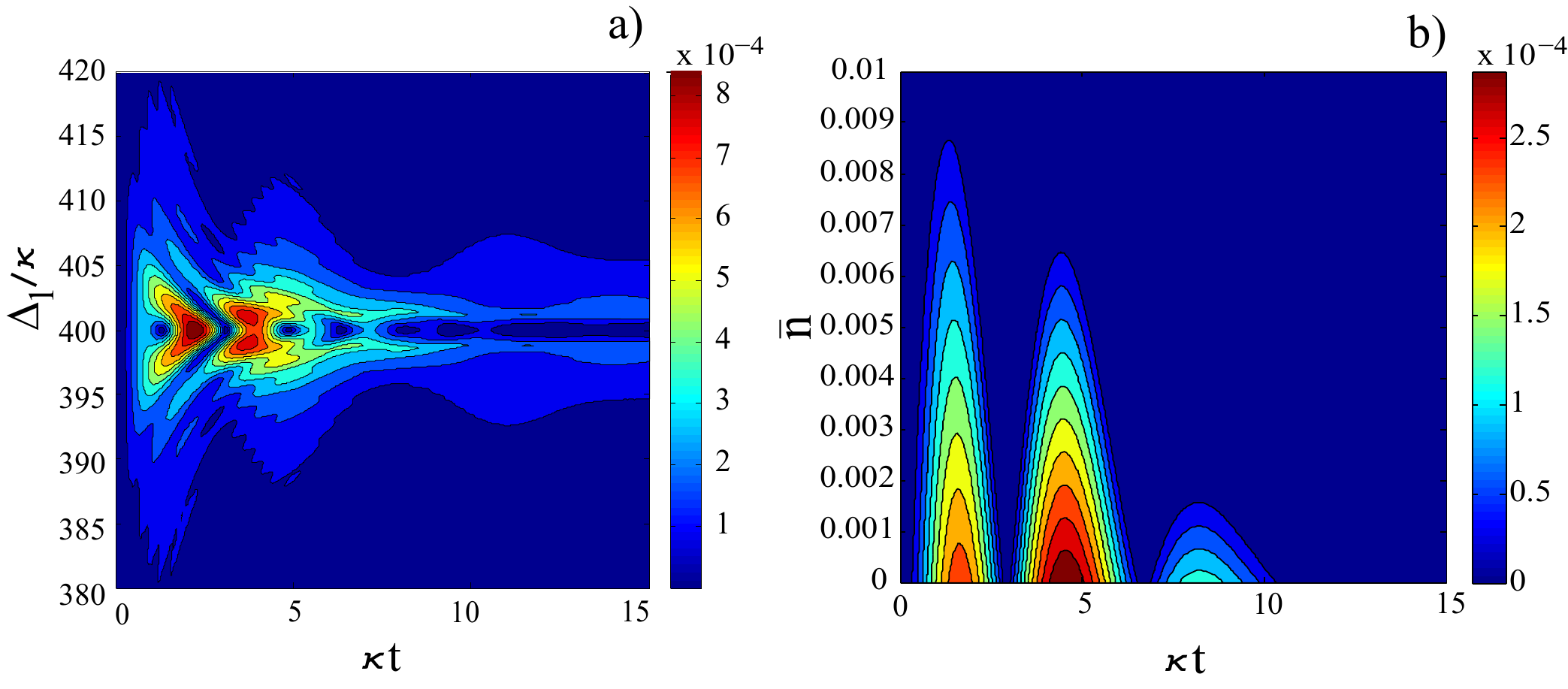}
\caption{(Color online) {\em Forward feed coupling with the full interaction.}\\The temporal entanglement, $E_{N}$ quantified by the logarithmic negativity  between intercavity phonons $b_{1},b_{2}$ of the optomechanical array as a function of (a) detuning of cavity 1 $\Delta_{1}$ and (b) thermal occupation number $\bar{n}$ where we have chosen $\kappa_{1,2,3}=\kappa$, $\mu_{1,2,3}/\kappa=0.01$, $\omega_{1}/\kappa=400$, $\Delta_{2,3}/\kappa=\omega_{2,3}/\kappa=400, g_{1,2,3}/\kappa=0.5$ and $\bar{n}=0 $.} 
\label{FFfulldelthermal}   
\end{figure}

When the three cavities are each detuned to the red sideband, we have found the intercavity entanglement is maximum when the detuning is resonant with the mechanical frequency of each cavity, i.e. $\Delta_{k}=\omega_{m_{k}}$ as illustrated in Fig.~\ref{FFfulldelthermal}(a) for $\Delta_{1}$. As previously the full linearised interaction Hamiltonian has been considered here, such that again the system is limited by different stability conditions on the blue and red sidebands as discussed in section~\ref{RRfull}. Hence the intercavity entanglement is found to exist only on the red sideband of the system, as the beam splitter part of the Hamiltonian induces the distribution of the intracavity optomechanical entanglement between the cavity modes, under large values of the optomechanical coupling strength, $g_{k}$, allowed by the stability conditions of the system. Thus the forward feed coupling between the optomechanical units also facilitates the distribution of intracavity entanglement over the intercavity modes. 

Fig.~\ref{FFfulldelthermal}(b) shows how the entanglement varies with an increase in mechanical thermal phonon number, $\bar{n}$. Clearly each mechanical mode of the coupled optomechanical array will be need to be as close as possible to its ground state to maintain the intercavity phonon phonon entanglement induced by the forward feed coupling between the optical modes.

\subsubsection{Unequal driving laser frequencies}
\label{bluered}
Analogous to the reversible  case we are now interested in the distribution of entanglement in optomechanical arrays given we have entanglement generated only in the source cavity. Again for simplicity we consider a chain of only two coupled cavities here, however our results can be extended for a large number of similarly coupled optomechanical systems.  We choose the source cavity to be exclusively on the blue sideband and the receiver cavity also to be only on the red sideband and have the same conditions for the cavity frequnecis and the detunings as in section~\ref{secRRbluered} such that the master equation in the RWA for the forward feed coupling now will be given as

\begin{eqnarray}
\frac{d\rho}{dt}&=& -\frac{i}{\hbar}[\acute{H}_I,\rho]+\sum_{k=1}^2\kappa_i{\cal D}[a_k]\rho +\mu_{k}(\bar{n}+1){\cal D}[b_{k}]\rho+\mu_{k}\bar{n}{\cal D}[b_{k}^\dagger]\rho \\ \nonumber
&&  + \sqrt{\kappa_1\kappa_2}\left ([a_1\rho, a_2^\dagger ] +[a_2, \rho a_1^\dagger ]\right )
\label{blueredME}
\end{eqnarray}
where making the rotating wave approximation at the frequencies, $\Delta_{1}=-\omega_{m}$ and $\Delta_{2}=\omega_{m}$ the time dependent coefficients in the coupling terms will be removed again. 
As before the squeezing interaction between the optical and mechanical modes in the source cavity generates entanglement which can be distributed along the chain in the array by taking the field out in the forward feed coupling and then swapping it into the mirror motion in the second cavity. Hence this system is capable of generating entanglement between the mechanical resonators, $b_{1}$ and $b_{2}$ even though the receiver cavity is explicitly on the red sideband. 

\begin{figure}[!h]
\centering
\includegraphics[scale=0.7]{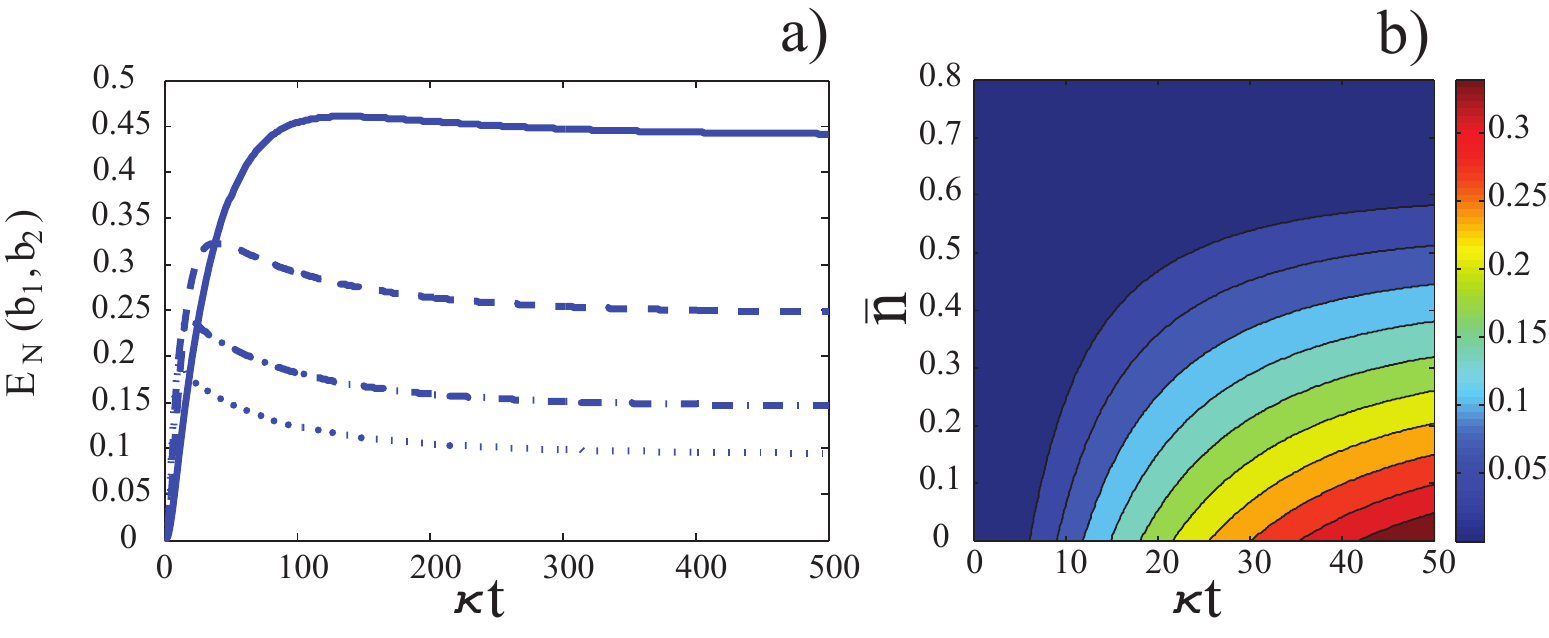}
\caption{(Color online) {\em Forward feed coupling with cavity 1 blue detuned and cavity 2 red detuned.}\\ The temporal evolution of entanglement, $ E_{N}$ quantified by the logarithmic negativity  between the intercavity phonons of two optomechanical cavities ($b_{1}$,$b_{2}$) where we have chosen $\kappa_{1,2} = \kappa$, $\mu_{1,2}/\kappa=0.01$, $g_{1}/\kappa=0.02$, (a) with $\bar{n}=0$ and different $g_{2}/\kappa$: 0.1 (solid line), 0.2 (dashed line), 0.3 (dashed dotted line) and 0.4 (dotted line) and (b) as a function of  thermal phonon number $\bar{n}$ with $g_{2}/\kappa=0.1$.}
\label{FFbluered}
\end{figure}  
We illustrate this in Fig.~\ref{FFbluered}(a), which shows how the entanglement varies between the two resonators as a function of time with all parameters in units of $\kappa$ for different optomechanical coupling strengths $g_{2}$ and for a fixed value of $g_{1}$.  As previously, the entanglement between $b_{1}$ and $b_{2}$,  becomes larger in the weak coupling regime of the receiver optomechanical system. In Fig.~\ref{FFbluered}(b) we show how the intercavity phonon-phonon entanglement varies in this coupling configuration under the effect of increasing the mechanical thermal phonon number. 
\begin{figure}[!h]
\centering
\includegraphics[scale=0.8]{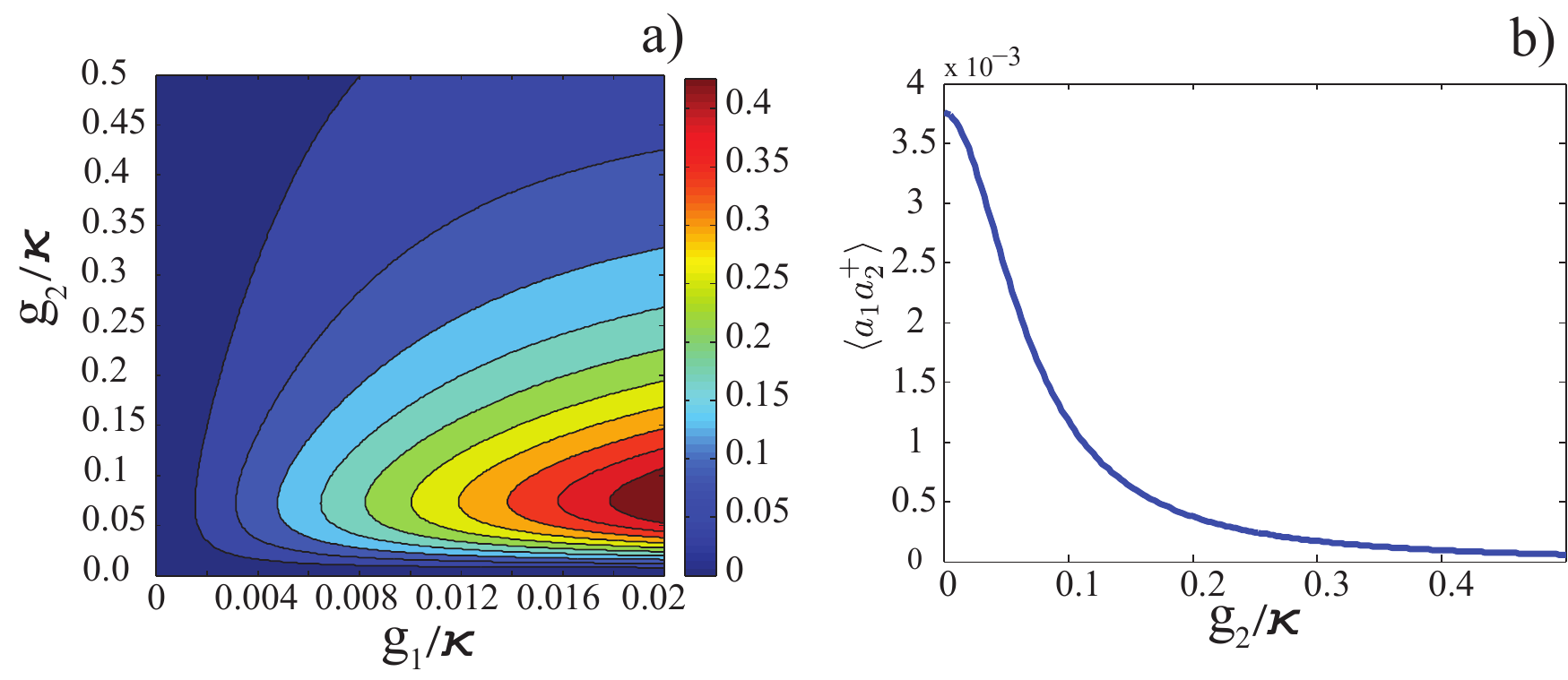}
\caption{(Color online) {\em Forward feed coupling with cavity 1 blue detuned and cavity 2 red detuned.}\\ (a) The steady state entanglement, $E_{N}$ quantified by the logarithmic negativity between the intercavity phonons of two optomechanical units ($b_{1}$,$b_{2}$) vs. $g_{1}/\kappa$ and $g_{2}/\kappa$ where we have chosen  $\kappa_{1,2} = \kappa$, $\mu_{1,2}/\kappa=0.01, $ and $\bar{n}=0 $ and (b) The steady state correlation $<a_{1}a_{2}^{\dagger}>$ as a function of $g_{2}/\kappa$ with the same parameters but with $g_{1}/\kappa=0.02$.}
\label{FFblueredSS}
\end{figure}

Here also we note that the intercavity entanglement quantified by the logarithmic negativity between the different modes is much larger compared to the case with the full linearised interaction Hamiltonian. We find the composite system has a steady state in this coupling configuration. Fig.~\ref{FFblueredSS}(a) shows the steady state entanglement between $b_{1}$ and $b_{2}$ as a function of different values of the optomechanical coupling strength of each cavity $g_{k}$. Fig.~\ref{FFblueredSS}(b) shows how the correlation between the optical modes $a_{1}$ and $a_{2}$ scales with $g_{2}$ for a fixed value of $g_{1}$. Similar to the reversible configuration, increasing $g_{2}$ leads to a much faster state swap between $a_{2}$ and $b_{2}$ such that the intercavity correlation between the optical modes, $<a_{1}a_{2}^{\dagger}>$, goes to zero along the $g_{2}$ axis. 

We note that the steady state results for the reversible and irreversible coupling configurations given in Figs.~\ref{SSRRbluered}(b) and ~\ref{FFblueredSS}(a) appear to be identical. The  effective coupling between the cascaded cavities in the reversible case is proportional to $\chi_{12}$ while in the irreversible case it is proportional to $\sqrt{\kappa_1\kappa_2}$. However the similarity is for the special case when the $\chi_{12}=\sqrt{\kappa_1\kappa_2}=1$ as chosen in Figs.~\ref{SSRRbluered}(b) and \ref{FFblueredSS}(a). The equations of motion resulting from each configuration are in fact different in the coupling terms and give different results for different values of the coupling parameters, $\chi_{12}$ and $\sqrt{\kappa_1\kappa_2}$. One can see this from the quantum stochastic differential equations for each coupling configuration given in the Appendix.

\section{Summary}
\label{summary}

We have analysed intercavity entanglement in an optomechanical array comprising of three coupled cavities, taking into account both a reversible coupling case via nearest neighbour evanescent coupling between the optical modes as well as a forward feed model realised through the cascaded systems approach. Further, for each coupling we have considered the effect of working with the full linearised interaction Hamiltonian (that is to say, without making the rotating wave approximation) such that both squeezing and beam splitter optomechanical interactions are present. The dynamics of such a coupled optomechanical array show that intracavity optomechanical entanglement generated independently in each cavity can be distributed pair wise between intercavity photons as well as phonons. The topology of the chain of oscillators considered in this paper comprising of reversible optomechanical coupling between light and mechanics, alongside irreversible forward feed coupling between optical ports, is quite distinct from earlier results on coupled oscillators. 

Moreover, for both irreversible and reversible configurations we also considered external coherent laser driving of each cavity such that the source cavity is explicitly driven on the blue sideband while the receiver cavity is only red detuned. In this choice of cavity-laser detunings, entanglement is only generated in the source cavity, while the receiver optical port simply swaps the entanglement from the field to the mechanics so that the mirrors $b_{1}$ and $b_{2}$ in the array become entangled. We find in this configuration, that the intercavity entanglement generated is much larger and exists in the steady state. It then varies with the optomechanical coupling strengths $g_{1}$ and $g_{2}$ within each optomechanical cavity. For the reversible coupling configuration, the steady state cross cavity phonon phonon entanglement increases with the reversible optical coupling strength $\chi_{12}$, while in a region of stability of each system as defined by the respective rotating wave approximations on the blue and red sideband of the optomechanical cavities. 

We also considered thermal effects on the entanglement generated in each case. It turns out the mechanical modes will need to be in the ground state to achieve intercavity phonon phonon entanglement but can also exhibit some degree of entanglement with a small non zero value of the phonon occupation number. While we illustrated the idea for only two cavities, this scheme can be extended to a chain of similarly coupled optomechanical units where the entanglement is generated only in the source unit but distributed along the entire chain due to state swap between the optics and mechanics. Here we have neglected delay effects in the forward feed case which means that delays must be less than the cavity decay times, which sets the time scale for the dynamics of the optomechanical array. Even with this assumption, the forward feed case enables the optical cavities to be much further apart --- up to one metre --- than the reversible coupling case.  Coupled optomechanical arrays could have wide applications as components of quantum repeaters and quantum memories required in quantum information processing.

\section{Appendix}

\subsection{Reversible coupling}

We have looked at the dynamics of the optomechanical arrays numerically, solving the master equation. In order to explain the steady state behaviour of the system we evaluate the steady state correlations between the intercavity phonons in this Appendix via the quantum stochastic differential equations for the coupled cavities. To keep the analytical calculations feasible we will only consider two cavities here coupled reversibly with the source driven on only the blue sideband and the receiver cavity on the red sideband, i.e. driving the system with different laser frequencies on each optomechanical unit.
Going into the interaction picture with respect to the driving laser fields, the collective system of the four oscillators can be described by the following closed set of coupled langevin equations,
\begin{eqnarray}
\frac{da_{1}}{dt} & = & -ig_{1}b_{1}^{\dagger}-\frac{\kappa_{1}}{2}a_{1}-i\chi_{12}a_{2}+\sqrt{\kappa_{1}}a_{in}\\ \label{a1}
\frac{da_{2}}{dt} & = & -ig_{2}b_{2}-\frac{\kappa_{2}}{2}a_{2}-i\chi_{12}a_{1}+\sqrt{\kappa_{2}}a_{in} \\ \label{a2}
\frac{db_{1}^{\dagger}}{dt} & = & ig_{1}a_{1}-\frac{\mu_{1}}{2}b^{\dagger}_{1}+\sqrt{\mu_{1}}b^{\dagger}_{1,in} \\
\label{b1qsde}
\frac{db_{2}}{dt} & = & -ig_{2}a_{2}-\frac{\mu_{2}}{2}b_{2}+\sqrt{\mu_{2}}b_{2,in} 
\label{b2qsde}
\end{eqnarray}
where $a_{in}$ is the vacuum input noise to the source cavity $a_{1}$, and $b_{k,in}$ is the noise for each mechanical resonator $b_{k}$. These linear equations can be solved analytically setting the time derivatives to zero, and employed to calculate all steady state correlations that exist between the different oscillators in the collective system. Eliminating both cavity fields we arrive at the following coupled quantum stochastic differential equations for the mechanical resonators $b_{1},b_{2}$ only,
\begin{eqnarray}
\frac{db_{1}}{dt}&=&-\frac{\gamma_{1}}{2}b_{1} +i\acute{\chi_{12}}b^{\dagger}_{2}+\eta_{1}^{*}a_{in}^{\dagger}+\sqrt{\mu_{1}}b_{1,in} \\ \label{b11}
\frac{db_{2}}{dt}&=&-\frac{\gamma_{2}}{2}b_{2} +i\acute{\chi_{12}}b^{\dagger}_{1}-{\eta_{2}}a_{in}+\sqrt{\mu_{2}}b_{2,in}
\label{b22}
\end{eqnarray}
where $\gamma_{1}=\mu_{1}-\Gamma_{1}$ and $\gamma_{2}=\mu_{2}+\Gamma_{2}$, such that
\begin{eqnarray}
\Gamma_{k}&=&\frac{4g_{k}^{2}\kappa_{j}}{\kappa_{k}\kappa_{j}+4\chi^{2}_{jk}} \\
\acute{\chi_{kj}}&=&\frac{4\chi_{kj}g_{k}g_{j}}{\kappa_{k}\kappa_{j}+4\chi^{2}_{kj}} \\
\eta_{k}&=&\frac{4\chi_{kj}g_{k}\sqrt{\kappa_{j}}+2i g_{k}\kappa_{j}\sqrt{\kappa_{k}}}{\kappa_{k}\kappa_{j}+4\chi^{2}_{kj}} 
\end{eqnarray}
where $k,j$ refer to the two reversibly coupled cavities in consideration.
To solve for the steady state correlation functions, we evaluate the expression,
\begin{eqnarray}
\frac{d(b_{2}b_{1})}{dt}=\left ( \frac{db_{2}}{dt}\right )b_{1} + b_{2}\left (\frac{db_{1}}{dt}\right )=0 
\label{b2b1t}
\end{eqnarray}
 which gives
 \begin{equation}
 <b_{2}b_{1}>=\frac{2i\acute{\chi_{12}}\left (1+<b_{1}^{\dagger}b_{1}> + <b_{2}^{\dagger}b_{2}>\right )}{\gamma_{1}+\gamma_{2}} 
\label{SSb2b1} 
\end{equation}

Eq.~\ref{SSb2b1} shows the presence of non zero steady state correlation and hence entanglement between the mechanical resonators $b_{1},b_{2}$. We note that increasing the reversible coupling strength $\chi_{12}$ will enhance the correlation and thus the entanglement as seen in Fig.~\ref{SSRRbluered}(b).

\subsection{Irreversible coupling}
Analogous to the reversible coupling, to examine the steady state behviour of the irreversible coupling we anlayse the quantum stochastic differential equations for two coupled cavities where the first is blue detuned while the second is red detuned. 
Going into the interaction picture with respect to the driving laser fields, the collective system of the four oscillators can be described by the following closed set of coupled langevin equations,
\begin{eqnarray}
\frac{da_{1}}{dt} & = & -ig_{1}b_{1}^{\dagger}-\frac{\kappa_{1}}{2}a_{1}+\sqrt{\kappa_{1}}a_{in}(t-\tau)\\ \label{a1FF}
\frac{da_{2}}{dt} & = & -ig_{2}b_{2}-\frac{\kappa_{2}}{2}a_{2}-\sqrt{\kappa_{1}\kappa_{2}}a_{1}(t-\tau)+\sqrt{\kappa_{2}}a_{in} \\ \label{a2FF}
\frac{db_{1}^{\dagger}}{dt} & = & ig_{1}a_{1}-\frac{\mu_{1}}{2}b^{\dagger}_{1}+\sqrt{\mu_{1}}b^{\dagger}_{1,in} \\
\label{b1qsdeFF}
\frac{db_{2}}{dt} & = & -ig_{2}a_{2}-\frac{\mu_{2}}{2}b_{2}+\sqrt{\mu_{2}}b_{2,in} 
\label{b2qsdeFF}
\end{eqnarray}
where $\tau$ is the time delay between the irreversibly coupled cavities. However here we neglect any time delays, and set $\tau=0$. 
As previously, eliminating both cavity fields we arrive at the following coupled quantum stochastic differential equations for the mechanical resonators $b_{1},b_{2}$ only,
\begin{eqnarray}
\frac{d b^{\dagger}_{1}}{dt}&=&-\frac{\tilde{\gamma_{1}}}{2}b_{1} +i\sqrt{\tilde{\Gamma_{1}}}a_{in}^{\dagger} +\sqrt{\mu_{1}}b^{\dagger}_{1,in} \\ \label{b11FF}
\frac{db_{2}}{dt}&=&-\frac{\tilde{\gamma_{2}}}{2}b_{2} +\sqrt{\tilde{\Gamma_{1}}\tilde{\Gamma_{2}}}b_{1}^{\dagger}+i\sqrt{\tilde{\Gamma_{2}}}a_{in}+\sqrt{\mu_{2}}b_{2,in}
\label{b22FF}
\end{eqnarray}
where $\tilde{\gamma_{1}}=\mu_{1}-\tilde{\Gamma_{1}}$, $\tilde{\gamma_{2}}=\mu_{2}+\tilde{\Gamma_{2}}$ and $\tilde{\Gamma_{k}}=\frac{4 g_{k}^{2}}{\kappa_{i}}$ with $k=1,2$. 
Again using Eq.~\ref{b2b1t} we arrive at the steady state correlation between $b_{1}$ and $b_{2}$,
\begin{equation}
<b_{2}b_{1}>=\frac{2 \sqrt{\tilde{\Gamma_{1}}\tilde{\Gamma_{2}}}}{\tilde{\gamma_{1}}+\tilde{\gamma_{2}}}<b_{1}^{\dagger}b_{1}>
\label{b2b1FF}
\end{equation}
which accounts for the intercavity steady state entanglement between the mechanical modes seen in Fig.~\ref{FFblueredSS}(a).

\section{Acknowledgements}
We would like to thank C. Joshi for useful discussions. We wish to acknowledge the support of the Australian Research Council through the Centre of Excellence for Engineered Quantum Systems. UA also acknowledges support from the University of Queensland Postdoctoral Research Fellowship and Grant.


\begin{references}


\bibitem{radpress1} T.J.Kippenberg, Rokhsari,H., Carmon,T., Scherer,A., Vahala,K.J. Phys. Rev. Lett. {\bf 95}, 033901 (2005). 

\bibitem{radpress2} O. Arcizet, P.-F. Cohadon, T. Briant, M. Pinard and A. Heidmann, Nature {\bf 444}, 71-75 (2006). 

\bibitem{radpress3} S. Gigan, H. R. B\"{o}hm, M. Paternostro, F. Blaser, G. Langer, J. B. Hertzberg, K. C. Schwab, D. B\"{a}uerle, M. Aspelmeyer and A. Zeilinger, Nature {\bf 444}, 67-71 (2006).

\bibitem{kippenbergreview} T. J. Kippenberg and  K. J. Vahala, Science {\bf 321},  1172 (2008).

\bibitem{APS-Physics} F. Marquardt and S. M. Girvin, Physics {\bf 2}, 40 (2009).

\bibitem{isart1} O. Romero-Isart, A.C. Pflanzer, M.L. Juan, R. Quidant, N. Kiesel, M. Aspelmeyer, J.I. Cirac, Phys. Rev. A {\bf 83}, 013803 (2011). 

\bibitem{isart2} O. Romero-Isart, Mathieu L. Juan, Romain Quidant, J.I. Cirac, New Journal of Physics {\bf 12}, 033105 (2010).

\bibitem{isart3} O. Romero-Isart, A.C. Pflanzer, F. Blaser, R. Kaltenbaek, N. Kiesel, M. Aspelmeyer and J.I.Cirac, Phys. Rev. Lett. {\bf 107}, 020405 (2011). 

\bibitem{chang} D.E. Chang, K.-K. Ni, O. Painter and H.J. Kimble, New J.Phys. {\bf 14}, 045002 (2012).

\bibitem{regal} D.E. Chang {\it et al.} Proceedings of the National Academy of Science of the United States of America, (PNAS), {\bf 107} no. 3 1005-1010 (2010).

\bibitem{corbitt} T. Corbitt {\it et al.} Phys. Rev. Lett. {\bf 98}, 150802 (2007). 

\bibitem{Schliesser} A. Schliesser, P. Del'Haye, N. Nooshi, K. J. Vahala and T.J. Kippenberg, Phys. Rev. Lett. {\bf 97}, 243905 (2006). 

\bibitem{eichenfield} Matt Eichenfield, Christopher P. Michail, Raviv Perahia and Oskar painter, Nature Photonics {\bf 1}, 416-422 (2007). 

\bibitem{markusstrongcoupling} S. Gr\"{o}blacher, K. Hammerer, M. R Vanner, M. Aspelmeyer, Nature, {\bf 460}, 724 (2009).

\bibitem{vacoptomechrate} M.L.Gorodetksy, Albert Schliesser, Georg Anetsberger, Samuel Deleglise, Tobias J. Kippenberg, Optics Express {\bf 18}, No. 22 23236 (2010).

\bibitem{Thompson} J. D. Thompson {\it et al} Nature {\bf 452}, 72-75 (2008).

\bibitem{verhagen} E. Verhagen, S.Delglise, S.Weis, A.Schliesser and T.J.Kippenberg, Nature {\bf 482}, 63 (2012). 

\bibitem{Marquardt} F. Marquardt, J. P. Chen, A. A. Clerk and S. M. Girvin, Phys. Rev. Lett. {\bf 99}, 093902 (2007).

\bibitem{Wilson-Rae} I. Wilson-Rae, N. Nooshi, W. Zwerger and T. J. Kippenberg, Phys. Rev. Lett. {\bf 99}, 093901 (2007). 

\bibitem{Genes} C. Genes, D. Vitali, P. Tombesi, S. Gigan, M. Aspelmeyer, Phys. Rev. A {\bf 77}, 033804 (2008).

 \bibitem{Connell} A. D. O Connell {it et al} Nature {\bf 464}, 08967 (2010).
 
 \bibitem{Kippenberg2011} R. Rivi�re, S. Del�glise, S. Weis, E. Gavartin, O. Arcizet, A. Schliesser, and T. J. Kippenberg, Phys. Rev. A, {\bf 83} 063835 (2011). 
 
 \bibitem{Teufel2011} J D Teufel, T Donner, Dale Li, J W Harlow, M S Allman, K Cicak, A J Sirois, J D Whittaker, K W Lehnert, R W Simmonds, Nature  {\bf 475}, 359 (2011). 

\bibitem{painter2011} Jasper Chan, T.P. Mayer Alegre, Amir H. Safavi-Naeini, Jeff T. Hill, Alex Krause, Simon Gr\"{o}blacher, Markus Aspelmeyer and Oskar Painter, Nature {\bf 478}, 89-92 (2011).

\bibitem{aspelmeyerreview} Markus Aspelmeyer {\it et al.,} J.Opt Soc. Am. B {\bf 27}, A189-A197 (2010). 

\bibitem{Sahar} S Basiri-Esfahani, U Akram and G J Milburn, New J. Phys. {\bf 14}  085017 (2012). 

\bibitem{Ludwig} Max Ludwig, Amir H. Safavi-Naeini, Oskar Painter and Florian Marquardt, Phys. Rev. Lett. {\bf 109}, 063601 (2012). 

\bibitem{amir} Amir H Safavi-Naeini and Oskar Painter New J. Phys. {\bf 13} 013017 (2011).

\bibitem{stannigel2} K. Stannigel,{\em et al.}, Phys. Rev. Lett. {\bf 109}, 013603 (2012).
 
\bibitem{muller} H. M. Ebhardt, H. Rehbein, R. Schnabel, K. Danzmann, and Y. Chen, Phys. Rev. Lett. {\bf 100}, 013601 (2008).

\bibitem{hartmann} M.J. Hartmann and M.B.Plenio, Phys Rev. Lett. {\bf 101}, 200503 (2008). 

\bibitem{vitali2} David Vitali, Stefano Mancini and Paolo Tombesi, J. of Physics A, {\bf 40}, 8055 (2007). 

\bibitem{vitali} D. Vitali {\it et al.,} Phys. Rev. Lett. {\bf 98}, 030405 (2007).

\bibitem{wipf} Christopher Wipf {\it et al.,} New J. Physics {\bf 10}, 095017 (2008). 

\bibitem{paternostro} L. Mazzola, M. Paternostro, Physical Review A {\bf 83}, 062335 (2011).

\bibitem{Heinrich} 	Georg Heinrich,  Max Ludwig, Jiang Qian, Bj�rn Kubala, and Florian Marquardt, Phys. Rev. Lett. {\bf 107}, 043603 (2011).

\bibitem{heinrichmarquardt}	Georg Heinrich and Florian Marquardt, Europhys. Lett. {\bf 93}, 18003 (2011).

\bibitem{stannigel} K. Stannigel, P. Rabl, P. Zoller, New J. Phys. {\bf 14}, 063014 (2012). 

\bibitem{joshi} C. Joshi, J. Larson, M. Jonson, E. Andersson and P. \"{O}hberg, Phys. Rev. A {\bf 85}, 033805 (2012) 

\bibitem{Car}H. J. Carmichael, Phys. Rev. Lett. {\bf 70}, 2273 (1993).

\bibitem{gardiner} C.W.Gardiner, Phys. Rev. Lett. {\bf 70}, 2269 (1993).

\bibitem{gard_zoller} C.W.Gardiner and P. Zoller, {\em Quantum Noise}, (Springer, Berlin 2004). 

\bibitem{eisertplenio} J. Eisert, M.B. Plenio, S. Bose, J. Hartley, Phys. Rev. Lett. {\bf 93}, 190402 (2004). 

\bibitem{vidal} G. Vidal and R.F.Werner, Phys. Rev. A {\bf 65}, 032314 (2002). 

\bibitem{optomechentanglement_vitali} C. Genes, A. Mari, D. Vitali, P. Tombesi, 'Quantum Effects in Optomechanical Systems', Adv. At. Mol. Opt. Phys. 57, 33-86 (2009). 







\end{references}
\end{document}